\documentclass{article}

\usepackage{arxiv}

\usepackage[utf8]{inputenc} 
\usepackage[T1]{fontenc}    
\usepackage{hyperref}       
\usepackage{url}            
\usepackage{booktabs}       
\usepackage{amsfonts}       
\usepackage{nicefrac}       
\usepackage{microtype}      
\usepackage{lipsum}		
\usepackage{graphicx}
\usepackage{natbib}
\usepackage{doi}
\usepackage{rotating}

\title{The Impact of Virtual Laboratories on Active  Learning \&  Engagement in Cybersecurity Distance Education}

\author{ \href{https://orcid.org/0000-0003-4071-4596}{\includegraphics[scale=0.06]{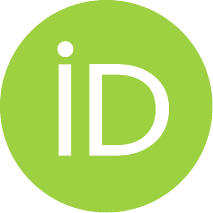}\hspace{1mm}Victor R.~Kebande}\thanks{Use footnote for providing further
		information about author (webpage, alternative
		address)---\emph{not} for acknowledging funding agencies.} \\
	Department of Computer Science\\
	Blekinge Institute of Technology\\
	Karlskrona, Sweden \\
	\texttt{victor.kebande@bth.se} \\
}

\renewcommand{\shorttitle}{\textit{arXiv} Template}

\hypersetup{
pdftitle={A template for the arxiv style},
pdfsubject={q-bio.NC, q-bio.QM},
pdfauthor={David S.~Hippocampus, Elias D.~Striatum},
pdfkeywords={First keyword, Second keyword, More},
}

\begin{document}
\maketitle

\begin{abstract}
	Virtual Laboratories ($VLabs$) have in the recent past become part and parcel of remote teaching in practical hands-on approaches, particularly in Cybersecurity distance courses. Their potential is meant to assist learners with  hands-on practical laboratory exercises irrespective of geographical location. Nevertheless, adopting $VLabs$ in didactic approaches in higher education has seen both merits and demerits.  Based on this premise, this study investigates the impact of $VLabs$ on Active Learning ($AL$) and engagement in cybersecurity distance education. A survey with a limited number of learners and educators who have had an experience with cybersecurity distance courses that leveraged $VLabs$ in their practical Lab assignment, was conducted at Blekinge Tekniska Högskola, Sweden, to assess the impact of $VLabs$ on $AL$ and engagement in Cybersecurity Distance Education.  $29\%$ and $73\%$ of the learners and educators, respectively responded to the survey administered remotely and with good internal consistency of questionnaires based on the Cronbalch Alpha; the results showed that learners and educators had a positive perception of using $VLabs$ to enhance $AL$ in cybersecurity distance education. The key concentration of the study was on $AL$  and engagement and problem-solving abilities when $VLabs$ are used. Both the learners and educators found the $VLabs$ to be engaging, interactive, and effective in improving their understanding of cybersecurity concepts. 
\end{abstract}


\section{Introduction}
Currently, cybersecurity has become an important and  critical area of concern that requires individuals with the necessary knowledge and skills to protect critical infrastructures from cyber-attacks. Higher Education Institutions play a crucial role in preparing students for careers in cybersecurity ~\citep{Jung2018}. Consequently, distance education has become an increasingly popular mode of delivering cybersecurity education due to its accessibility and flexibility. However, distance education faces challenges in terms of student engagement and learning, especially in the field of cybersecurity where hands-on practical experience is crucial ~\cite{khan2017active,fazza2021student}.

The need to measure Active Learning ($AL$) and the use of Virtual Laboratories ($VLabs$) are two approaches that have shown the potential in enhancing student learning in distance education. $AL$ is a student-centered approach that emphasizes the importance of student participation and engagement in the learning process ~\cite{lumpkin2015student}. $VLabs$, on the other hand, provide students with simulated hands-on experience in a controlled environment, thus enhancing their practical skills and understanding ~\cite{chiu1999benefits}.

Despite the potential benefits of $AL$ and engagement and the use of $VLabs$, there is a lack of an effective approach of ascertaining the  impact posed by $VLabs$ on $AL$ and engagement in cybersecurity distance education at the time of writing this article. This is a considerable challenge as it makes it difficult to assess effectiveness of $VLabs$  in enhancing and promoting student learning. Consequently, assessing the suitability  of $VLabs$ in cybersecurity distance education is crucial to inform the development of pedagogical approaches that can improve student outcomes based on their supposed engagement.

The aim of this study is to assess the impact of $VLabs$ on $AL$ and engagement ($AE$) in cybersecurity distance education. While the study explores diverse aspects like $AE$ in Course Concepts ($AE$), Active Learning and Engagement and Problem Solving ($AE$ \& $PS$), difficult Course Concepts (CC), help in Practicing Course Concepts (PCC) and Critical Thinking (CT), and Comparison to Traditional Learning (TL). However, the focus of this study narrows down to $AE$  and $AE$ \& $PS$. The contribution of this study can be summarised as follows:

\begin{itemize}
    \item  \textbf{Impact of $VLabs$ on $AL$ \& engagement ($AE$) and Problem-Solving ($PS$):} The study aims to investigate the impact of Virtual Laboratories ($VLabs$) on active learning ($AL$) and engagement ($AE$), as well as problem-solving ($PS$) in the context of cybersecurity distance education. By exploring the potential benefits of incorporating $VLabs$ into the learning environment, the study seeks to shed light on the effectiveness of this innovative approach in promoting student learning outcomes.
    \item \textbf{Statistical significance:}  The study conducts an independent sample test to see if there exist an statistical significance between $AE$ and $PS$ when using  $VLabs$ in cybersecurity distance education. In addition, the study examines if there exist any relationship between $AE$  and  $PS$ when using  $VLabs$ in cybersecurity distance education 
        \item  \textbf{Comparison with existing studies:} This compares the proposed concept with previous and existing studies as a measure of validating and reinforcing the findings and identifying further gaps and inconsistencies if available.
    
\end{itemize}


This study is relevant as it addresses a significant gap in the literature on the use of $AL$ and $VLabs$ in cybersecurity distance education. In addition this study is important because it provide insights into how to design effective instructional strategies and pedagogical approaches in cybersecurity distance education and enhance student outcomes.  Evaluating the impact of $AL$ when leveraging $VLabs$ proves to be an essential approach for instructors and researchers in the field of cybersecurity education, enabling them to determine the effectiveness of these approaches in enhancing student learning outcomes. 


The remainder of this paper is structured as follows: Section II gives the Background, which is followed by the $VLab$ Process Model in Section III. The methodology is given in  Section IV. The Results are discussed in Section V. A comparison with existing studies is given in Section VI while the discussions of the proposed concepts are given in Section VII. Limitations and Future studies are given in Section VII.  Concluding Remarks are discussed in Section VIII.

\section{Background}
\subsection{ Theoretical Framework}
\noindent It is pertinent to ascertain that the targeted learners in the current exercise are exposed to $AL$ and engagement strategies. As a result, this study's theoretical framework is inclined towards constructivism, which puts the emphasis for learners to actively construct their understanding of the subject based on exposure of some important experiences~\cite{jonassen1999activity}. In addition, learning in this context is viewed as a dynamic process that involves the learner's active engagement to cybersecurity education ~\cite{kahu2013framing}. 

\begin{figure}[htbp]
  \centering
  \includegraphics[width=0.6\textwidth]{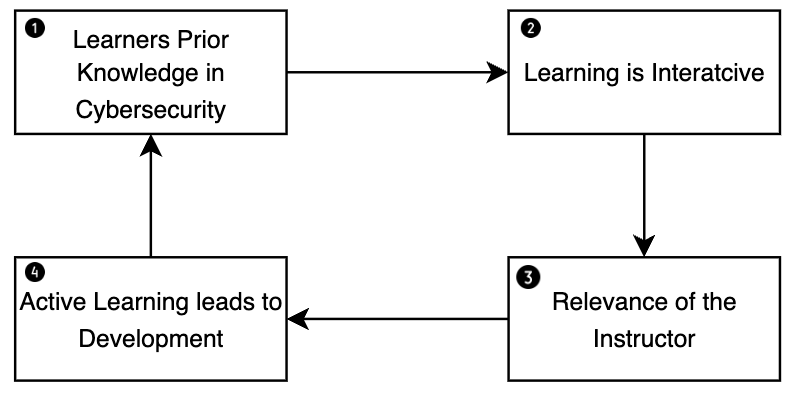}
  \caption{Theoretical Framework setting the scene of this study based on principles of constructivism theory}
  \label{fig:image}
\end{figure}

By utilizing $VLabs$ as a means of assessing active learning ($AL$), this approach aligns with the principles of constructivism theory, owing to the fact that,  it provides the learners with an opportunity to have hands-on activities and experiences that in the long-run promotes the construction of their knowledge. In addition, $VLabs$ are positioned to provide learners with a collaborative and active engagement environment that is based on an interaction with educators and other peers-which is essential in constructivism theory because it  allows the development of critical and problem-solving skills in cybersecurity education. Figure \ref{fig:image} shows the role of constructivism theory for cybersecurity which also aligns with the pedagogy approaches as highlighted by Vygotsky ~\cite{prawat2002dewey}.

Step 1  of Figure \ref{fig:image} highlights the problem that learners could encounter as a result of having or not having cybersecurity knowledge, whereas Step 2 concentrates on the effects of interactive learning and the relevance of instructors is shown in Step 3, which ultimately leads to the learner's development as is shown in Step 4. While the theoretical framework shown in Figure \ref{fig:image} is grounded in constructivism, its emphasis is positioned to active learner engagement and the construction of knowledge based on meaningful experiences in the context of; active learner engagement and knowledge construction could include hands-on activities, problem-solving tasks, real-world case studies, collaborative discussions, and interactive simulations.

\subsection{ Related Work}
\noindent To highlight the significance of distance education in the field of cybersecurity and the specific challenges it presents, it is observed that  distance education has become a popular mode of delivering education, especially for cybersecurity due to its accessibility and flexibility. However, teaching cybersecurity in a distance education setting poses several challenges, such as ensuring students have access to hands-on learning experiences and resources that simulate real-world scenarios. $VLabs$ offer a solution to these challenges by allowing students to access and work on simulated cyber attacks and defenses in a controlled environment. Studies have shown that $AL$ is an effective teaching method in higher education, with research indicating that it can lead to improved student engagement, learning outcomes, and critical thinking skills ~\cite{Bonwell1991}, ~\cite{Freeman2014}. $AL$ has also been found to be particularly effective in science, technology, engineering, and mathematics (STEM) fields, including cybersecurity ~\cite{Chuang2018}. $VLabs$ have been used in higher education for several decades and have been shown to be effective in providing students with hands-on learning experiences, particularly in STEM fields ~\cite{Bjarnason2010}. $VLabs$ are especially useful in distance education as they allow students to access and work on simulated scenarios, providing a controlled environment for learning that would not be possible in the real-world (~\cite{Huang2019}.Studies have also shown that $VLabs$ can be used effectively in cybersecurity education, providing students with opportunities to work on simulated cyber attacks and defenses in a controlled environment ~\cite{Kapetanakis2016}. In the context of cybersecurity distance education, $VLabs$ can provide students with hands-on experiences that would not be possible in a traditional classroom setting ~\cite{Tsai2020}. However, there is limited research on the impact of AL and engagement activities centered in  $VLabs$ with particular interest to student learning in cybersecurity distance education ~\cite{Jung2018}. Therefore, there is a need for approaches that can assess the effectiveness of this approach to teaching and learning. 

\section{VLabs process Model  }

\begin{figure*}[htbp]
  \centering
  \includegraphics[width=0.7\textwidth]{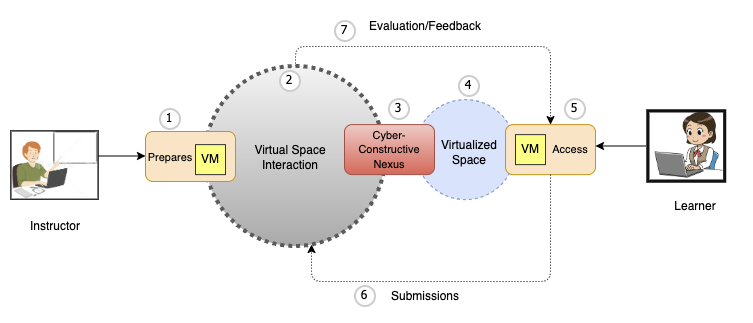}
  \caption{Sequences in the  $VLabs$ processes ~\cite{momen2023constructivism}.}
  \label{fig:image2}
\end{figure*}

Virtual Laboratories ($VLabs$) sequences provides an integral process that allows learners to be able to have a hands-on approach, with an interactive environment that is supported by remote execution of practical assignment over emulations of computers in cybersecurity distance education. $VLabs$ can support multiple task, share data, and real-time collaborative execution of hands-on practical assignments. Figure \ref{fig:image2} shows  $Vlab$ processes that are $labelled$ $1$ to $5$. The Educator Module (EM) $labeled$ $1$ allows the instructor to create the $VLab$ exercises. It serves as as the genesis of the preparation of course materials and instructions to learners which is done in the step $labeled$ $2$ with the aid of provisioning a $Virtual$ $Machine$ (VM) over a Virtual Space Interaction (VSI). A $VM$ is a software emulation of a physical computer that allows multiple Operating Systems (MOS) to run simultaneously on a single computer. Once the $VLab$ exercise is ready, learners are able to receive instructions and guidelines from the educators regarding the objectives, tasks, and resources available in the $VLabs$. 

This interaction enables the learners to be able to work with a logical abstraction of a computer and allows them to inquire, seek clarification and execute tasks and receive real-time feedback as they engage with the $VLab$. A Cyber-Constructive Nexus (CCN) in the step $labeled$ $3$ provides a link between the VSI and the virtualized pace ($labeled$ $4$). Through the learner's module in the step $labeled$ $5$, learners interact with the $VLab$ through a VM. They access the VLab platform, where they can perform hands-on activities, experiments, and simulations related to cybersecurity concepts and practices in readiness for assessment. This hands-on experience allows learners to apply theoretical knowledge, develop practical skills, and gain a deeper understanding of the subject matter. Throughout the learner's engagement with the $VLabs$, there is a continuous exchange of instructions and communication between the educator and the learner in the step $labeled$ $6$ and $6$, however, not in real-time in some cases. 

\begin{figure}[htbp]
    \centering
    \includegraphics[width=0.6\textwidth]{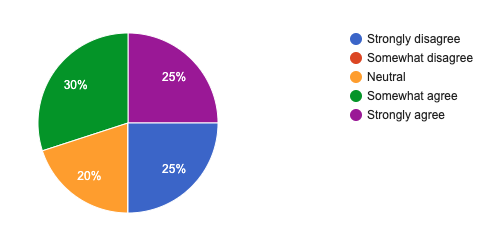}
    \caption{Proportion of learners that feel Active   learning and engagement has increased interest in the course}
    \label{fig:myimage}
\end{figure}

\section{Methodology}
\noindent In order to assess the impact of $VLabs$ in cybersecurity distance education, a three-fold methodology has been leveraged encompasing, sampling and face validation, assessment and testing. The three-fold approach has been used to assess the learners and educators outcomes based on purposive sampling where a list of learners  who had experienced the use of $VLabs$ was considered, which included face validity approaches that allowed cybersecurity experts to give expert reviews based on the prevailing research question. This study is  cross-sectional, owing to its collection of data at a single point in time from different individuals and groups in order to examine relationships or differences. Also, the study leverages a quasi-experimental approach with pre-test that have been used to test the significance. Other subprocesses that have been used in the study are explained in the subsequent sections.

  \begin{table}[htbp]
\centering
\caption{Learners' and Educators' Questionaire  (Cronbach Alpha)}

\label{tab:mytablepp}
\begin{tabular}{@{}cc@{}}
\toprule
\hline
\textbf{Target} & \textbf{Cronbach Alpha }\\
\hline
\midrule
Learners Questionnaire & 0.791 \\
\hline
Educator's Questionnaire & 0.847 \\
\hline
\end{tabular}
\end{table}

\subsection{Context and Participants Selection}
\noindent The participants consisting of learners, were recruited from the Department of Computer Science at Blekinge Institute of Technology, Sweden. A random and voluntary survey with questions shown in Table \ref{tab:mytablevv} was passed to learners  who currently or previously had leveraged $VLabs$ in Cybersecurity Distance Education in the past 3 years. The participants constituting educators who have had experiences with $VLabs$ were non-randomly recruited using purposive sampling recruited from BTH, Sweden, Malmö University, Sweden, Curtin University, Australia, Halmstad University, Sweden, and Zayed University, UAE. A survey with question shown in Table \ref{tab:mytableww} was passed to the educators in order to assess theor perceptions on use of $VLabs$. In addition, there was no significance of identifying the participant's age, gender and ethnicity among learners and educators in this study.

\begin{figure}[htbp]
    \centering
    \includegraphics[width=0.6\textwidth]{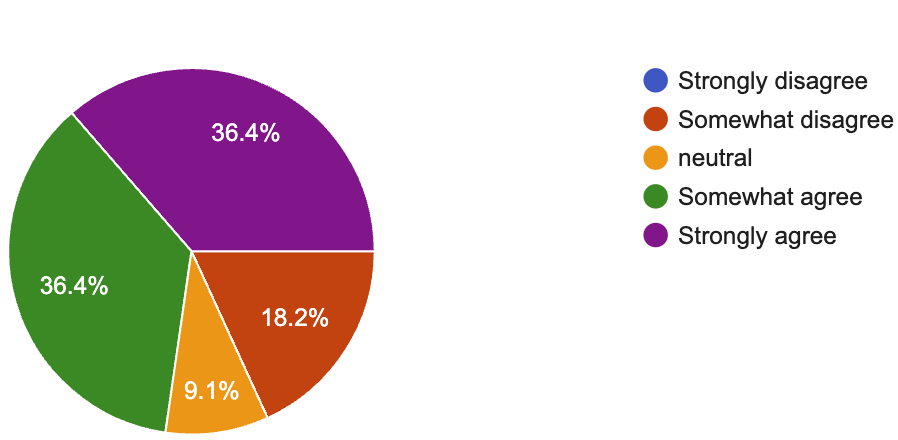}
    \caption{Proportion of Educators that feel Active  learning and engagement has increased interest in the course}
    \label{fig:myimage11}
\end{figure}

\subsection{Test Surveys}
 \noindent The survey aimed to assess $VLabs$' impact on $AL$ and engagement in Cybersecurity Distance Education. A relatively, similar set of questions was administered to both the learners and educators. Six (6 ) items were assessed using a 5-point Likert scale ranging from "Strongly Disagree-1" to "Strongly Agree-5".

\begin{table*}[htbp]
  \small 
\centering
\caption{Administered Learners Questionnaire}

\label{tab:mytablevv}
\begin{tabular}{@{}cc@{}}
\hline
\textbf{Survey question} & \textbf{Scale } \\
\hline
\noindent Do you feel that the use of $VLabs$ have helped you understand difficult concepts in your course? & 1-5  \\
\noindent Do you feel that the use of $VLabs$ have have improved your AL \& engagement and interest in the course material?   & 1-5 \\

\noindent Is the use of $VLabs$ in distance courses most appropriate when compared to traditional (in-person) laboratories?  & 1-5 \\
\noindent Do you feel that $VLabs$ have allowed you to apply the course concepts in practical setting? & 1-5 \\
\noindent Do you feel that using $VLabs$ have improved your thinking skills? & 1-5 \\
\noindent Do you feel that taking practicals virtually allows you to actively learn \& engage in problem-solving, discussions, and \\ hands-on activities well as compared to traditional approaches? & 1-5 \\

\hline

\end{tabular}
\end{table*}

\begin{table*}[htbp]
  \small 
\centering
\caption{Administered Educators Questionnaire}

\label{tab:mytableww}
\begin{tabular}{@{}cc@{}}
\hline
\textbf{Survey question} & \textbf{Scale } \\
\hline
\noindent Have you used $VLabs$ in giving cybersecurity/online/distance courses? & Y/N  \\
\noindent Do you feel that giving $VLabs$ has helped distance students to understand the difficult concepts in the course?   & 1-5 \\

\noindent Do you feel that $VLabs$ have improved distance  students AL \& engagement and interest in the course materials?  & 1-5 \\
\noindent Is the use of $VLabs$ in distance courses most appropriate when compared to traditional (In-person) laboratories? & 1-5 \\

\noindent Do you feel that giving $VLabs$ to distance students have allowed students \\to apply the course concepts in practical setting? & 1-5 \\

\noindent Do you feel that giving $VLabs$ have improved student thinking skills? & 1-5 \\
\noindent Do you feel that distance students taking practicals virtually do actively learn engage in problem-solving, \\discussions, and hands-on activities well as compared to traditional approaches  & 1-5 .\\
\hline

\end{tabular}
\end{table*}

 There were four subscales namely: Course concepts, active engagement, comparison to traditional approaches, and critical thinking skills. Two (2) items  were related to the general learning course concepts, while two (2) items had a relationship to $AL$ and engagement and one (1) item was inclined to the preference of $VLabs$ to traditional approaches. In addition to this, three (3) items focused on applying the skills based on the constructivism theory ~\cite{jonassen1999activity}, as was mentioned in Section II of this paper. Among the 70 learners who were randomly selected to participate in the survey, 20 consented to participate in this activity,  and observations were made from the 20 responses, representing approximately a response rate of 29\%. Consequently, 11 out of 15 educators responded with a response rate of approximately 73\%. From the 6 items that were assessed for the learners and educators, a Cronbach's Alpha of 0.791 and 0.847 were realized respectively as is shown in  Table \ref{tab:mytablepp}, which indicates a good level of internal consistency and relatively high reliability given that Cronbach's alpha coefficient of 0.7 or higher is considered acceptable. The maximum possible score that could be achieved in the assessment was 30, a score above the median is regarded as a good engagement with $VLabs$ among learners which implies learners appreciate the concepts in $VLabs$.

  \begin{table*}[htbp]
\centering
\caption{Descriptive statistics for the Educators  }

\label{tab:mytablessxx}
\begin{tabular}{@{}ccccccc@{}}
\toprule
\hline
{Dimension}& \textbf{AE} & \textbf{$AE$ \&$PS$ } & \textbf{CC} & \textbf{TL } & \textbf{PCC} & \textbf{CT}\\
\hline
\midrule
 N& 11 & 11& 11& 11& 11& 11\\
\hline
 Mean& 3.91 & 3.27& 3.91& 3.45& 4.18& 4.27\\
\hline
 Median& 4.00 & 3.00& 4.00& 4.00& 5.00& 4.00\\
\hline
 SD & 1.044 & 1.421& 1.136& 1.128& 1.250& .786\\
\hline
 Variance & 1.091 & 2.018& 1.291& 1.273& 1.564& .618\\
\hline
 Skewness & -1.074 & -.074& -.789& -.118& -1.912& -.574\\
\hline
 Kurtosis & .581 & -1.384& -.546& -1.306& 3.867& -.967\\
\hline
\end{tabular}
\end{table*}

 \begin{table*}[htbp]
\centering
\caption{Descriptive statistics for the Learners  }

\label{tab:mytablex}
\begin{tabular}{@{}ccccccc@{}}
\toprule
\hline
{Dimension} & \textbf{$AE$} & \textbf{$AE$ \&$PS$ } & \textbf{CC } & \textbf{TL } & \textbf{PCC } & \textbf{CT }\\
\hline
\midrule
 N& 20 & 20& 20& 20& 20& 20\\
\hline
 Mean& 3.30 & 3.05& 3.25& 3.40& 3.25& 4.27\\
\hline
 Median& 4.00 & 3.00& 4.00& 3.50& 5.00& 4.00\\
\hline
 SD& 1.525 & 1.538& 1.517& 1.603& 1.482& 1.603\\
\hline
 Variance& 2.326 & 2.366& 2.303& 2.568& 2.197& 2.568\\
\hline
Skewness& -.566 & -.382& -.370& -.396& -.263& -.713\\
\hline

Kurtosis& -1.094 & -1.512& -1.36& -1.306& -1.476& -.1.082\\
\hline

\end{tabular}
\end{table*}

\subsection{Data Collection }

The survey encompassing questionnaires shown in Tables \ref{tab:mytablevv} and \ref{tab:mytableww} respectively was distributed to learners and educators via university student and staff email between February and March 2023. In order to recruit participants, the survey was created using Google Forms and it was sent using as a link to participants twice in a span of one month. Informed consent that provided information regarding the survey stated that  the study was voluntary and anonymous and the responses were kept confidential. It also stated that the participants were free to withdraw at will. Furthermore, the consent also highlighted that the outcome/result from the survey was explicitly to inform the impact of $VLabs$ on $AL$ and engagement in cybersecurity distance education. Furthermore, it is worth mentioning, even though the study was anonymous, only the study investigator had access to the participant's data, and this data was securely stored.

\subsection{Data Analysis}
IBM Statistical Package for Social Sciences (SPSS) V28 was used to analyze data after data had been cleaned. A probability of <0.05 was considered to be significant for the test conducted (alpha=0.05). This  involved descriptive statistics that allowed calculation of the central tendencies; Mean, median, Standard deviation,variance, Skewness, Kurtosis and analysis of descriptive variables for both the learners and educators in Cybersecurity distance education as is shown in Tables \ref{tab:mytablessxx} and \ref{tab:mytablex} respectively. The considered aspects in this context included $AL$ and Engagement in Course Concepts ($AE$), Active Learning and Engagement and Problem Solving ($AE$ \& $PS$), difficult Course Concepts (CC), help in Practicing Course Concepts (PCC) and Critical Thinking (CT), and Comparison to Traditional Learning (TL), however, the focus of this study was inclined towards  $AE$  and $AE$ \& $PS$. In addition, inferential statistics were used to test hypotheses and determine statistical significance. The results of the data analysis will provide insights into the perceptions of $AL$ and engagement among cybersecurity distance learners and inform future research and practice in this field.

\section{Results}

\subsection{Demographics}
Twenty learners (20) learners and 11 educators participated in this study which amounted 29\% and 73\% respectively. Learners were selected randomly, from the Network and System Security Course, which is part of Msc Telecommunication Programme at Blekinge Institute of Technology, BTH, Sweden. Of the participants, 63\% of the educators responded to have used $VLabs$ in giving cybersecurity distance courses, 9\% of the educators were not sure and 27\% had not used $VLabs$. In the $AL$ and engagement on the course, 36.4\% of the educators and 25\%  of the learners had a strong perception about the effect of $VLabs$ as  is shown in Figure \ref{fig:myimage} and Figure \ref{fig:myimage11} respectively. The significant aspects that were observed aspects like $AL$ and Engagement in Course Concepts ($AE$), Active Learning and Engagement and Problem Solving ($AE$ \& $PS$), difficult Course Concepts (CC), help in Practicing Course Concepts (PCC) and Critical Thinking (CT), and Comparison to Traditional Learning (TL), however, the focus of this study narrows down to $AE$  and $AE$ \& $PS$ which are discussed in Sections 5.2 and 5.3 respectively. Although all the items ($AE$,\& $PS$,CC,PCC,CT and TL are studied and included in the study as part of the responses from the learners and educators, it is worth mentioning that, the motive of exploring this concepts is that, they have been used to set the stage for the specific aspects that forms the major part of the study of this paper ($AE$ \& $PS$). It should also be noted that the inclusion of these concepts do not influence the outcome of the study. This narrowed focus allows for a more in-depth analysis of the effects of VLabs on these specific dimensions of learning and engagement in the context of cybersecurity education.


\subsection{ AL and   Engagement on the course }

In examining the learner's and educator's impact on engagement and interest of the course when $VLabs$ are used showed the following was observed; 25\% of the learners strongly agreed that the use of $VLabs$ have had a positive impact on the course, which included improved engagement and interest in the course material. Also, 25\% strongly disagreed, 30\% somewhat agreed to have seen the benefits and 20\% were neutral. On the same note, 36.4\% of the educators agreed that using $VLabs$ has had a huge impact on student engagement and interest to the course with 36.4\% somewhat agreeing,18.2\% somewhat disagreed, with 9.1\% staying neutral. The responses from learners and educators on the use of $VLabs$  and Engagement on the course  reported (M=3.30, SD=1.525 and 3.91 and 1.044 respectively)

\subsection{ AL and Problem-Solving}

Descriptive statistics for the impact of $VLabs$ on AL and engagement was meant to assess whether $VLabs$ actively helps the students in their problem-solving skills. 30\% of the learners said that the use of $VLabs$ had not allowed them to actively engage in problem-solving, having key study discussions and helping them with hands-on activities. 40\% somewhat agreed that the use of $VLabs$ had helped. Only 15\% of the learners attributed $VLabs$ to have strongly helped them with their active engagement in problem-solving abilities. 10\%, however, stayed neutral with 5\% somewhat disagreeing.

Likewise, 27.3\% of the educators strongly agreed that $VLabs$ had improved active engagement and problem-solving abilities for learners with 18.2\% of the educators somewhat agreeing to this. However, 27.3\% of the educators somewhat disagreed with this and 9.1\% strongly disagreed with this. 18.2\% of the educators chose to be neutral. The responses from learners and educators on the use of $VLabs$  to assess the ability of problem solving in the  course  reported (M=3.05, SD=1.538 and 3.27 and 1.421, respectively)







\subsection{Independent T-test Analysis }
In order to check the statistical significance between $AL$, engagement and problem-solving among the learners and educators responses, an independent T-test was conducted in order to analyze the data from the two groups as is shown in Table 6. Independent t-test was chosen as the appropriate statistical test that could compare the means of the variables in those groups. The hypothesis that guided this analysis was as follows:

 \begin{sidewaystable}[htbp]  
  \small
  \centering
  \caption{Independent Sample Test}
  \label{tab:mytablexvx}
  \begin{tabular}{@{}cccccccccccc@{}}
    \toprule
    \hline
    {Dimension} & {Var} & \textbf{F} & \textbf{Sig} & \textbf{t} & \textbf{df} & \textbf{1-sp} & \textbf{2-sp} & \textbf{MD} & \textbf{S.E D} & \textbf{Lw} & \textbf{Up} \\
    \hline
    \midrule
    Active Learning in CM & EV Assm & 2.917 & 0.131 & -7.514 & 7 & <.001 & <.001 & <-3.667 & <0.488 & <-4.820 & <-2.513 \\
    \hline
    - & EV Not Assm & - & - & -11.000 & 5.000 & <.001 & <.001 & <-3.667 & <.333 & <-4.524 & <-2.810 \\
    \hline
  \end{tabular}
\end{sidewaystable}

\textit{\textbf{Ho:} There is no difference in the means of active learning and engagement, and problem solving between the Learners and Educators}

The purpose of this hypothesis was to determine whether there exist any significant or statistical differences in the levels of  $AL$ and engagement and problem-solving between the two groups. The independent T-test analysis mainly covered two independent variables from the learner's group as is shown in \ref{tab:mytablexvx}. The dimension showing Active Learning in CM, shows the $F-Value$ of 2.917 and a corresponding p-value of 0.131. The t-value shows -7.514 and a degree of freedom (dof) of 7.

The one-sided p-value (1-sp) and two-sided $p-value$ (2-sp) are both less than 0.001. In addition, the Mean Difference (MD) was calculated as -3.667 with s Standard Error (S.E) of 0.488. The lower (Lw) and Upper (Up) confidence intervals are given as -4.524 and -2.810 respectively. The independent t-test gives $F-test$ based on Levene's test of equality of variances was used to check if the variance is equal. Since the study do  not have an alternative hypothesis, it was observed that our significance for p-value for $F-test$ is less than $alpha$= $(0.05)$, indicating that the variance are not assumed to be equal since the p-value is <0.05, the groups  that had a response on $AL$ and problem-solving are not assumed to be the same.

The results from the independent-t-test provide differences in terms of $AL$, engagement, and problem-solving among learners and educators. The significant negative $t$ value shows that there exist statistical significant differences of the variables from the two groups. Therefore having a p-value  $0.01 <0.05$ than the initially chosen significance level of 0.05 show that the null hypothesis, Ho falls on the regions of rejection.

From the findings, conclusions can be made that there exist significant differences in the means of $AL$ and engagement, and problem-solving among learners and educators. This shows that the learners and educators have distinct levels of active learning, engagement, and problem-solving as far as cybersecurity distance courses are concerned.


\subsection{Correlation  Analysis}

A correlation analysis meant to examine the relationship between $AL$  and engagement, and problem-solving among learners and educators was conducted as is shown in  Table \ref{tab:mytablewoo} and \ref{tab:mytableccv} respectively.

Table \ref{tab:mytablewoo} shows the correlation coefficients between $AL$ and engagement ($AE$) and problem-solving ($AE$ \& $PS$) for the learners. The Pearson correlation coefficient $(r)$ was used to determine the strength and direction of the relationship, while the p-value was used to assess the statistical significance of the correlation. The Pearson correlation coefficient between $AL$ and engagement is 0.891. This indicates that there is a strong positive correlation among these variables in learners with  $(r = 0.891, p < 0.05)$ as is shown in Table  \ref{tab:mytablewoo}. The correlation is statistically significant $(p < 0.001)$, which indicates that an increase in $AL$ so does the level of engagement among the learners. Also, the correlation coefficient for $AE$ \& $PS$ is 0.891 which also indicates a strong positive correlation. Consequently, it is also statistically significant at $(p < 0.001)$ showing that an increase of $AL$ triggers the increase the level of engagement and problem-solving skills among learners.

\begin{table*}

\centering
\caption{Learner's correlations' $VLabs$ engagement and Active Engagement and problem solving}

\label{tab:mytablewoo}
\begin{tabular}{@{}ccc@{}}
\hline
\textbf{Correlations outcomes} & \textbf{$AE$ } &  \textbf{$AE$ \&$PS$ }\\
\hline
\textbf{$AE$} &-  &- \\
Pearson r & 1  & 0.891\\

Sig. (two-tailed)&  & <.001\\
N&20  & 20\\
\hline
\textbf{$AE$ \&$PS$} &-   &- \\
Pearson r & 0.891  & 1\\
Sig. (two-tailed)& <.001 &-\\
N&20  & 20\\
\hline

\multicolumn{3}{p{251pt}}{Correlation is significant at 0.001 level (two-tailed) }\\
\end{tabular}
\end{table*}

\begin{table*}
\centering
\caption{Educator's correlations' AL \& engagement and Active Engagement and problem solving}

\label{tab:mytableccv}
\begin{tabular}{@{}ccc@{}}
\hline
\textbf{Correlations outcomes} & \textbf{$AE$ } & \textbf{$AE$ \&$PS$  }\\
\hline
\textbf{$AE$} &-  &- \\
\centering Pearson r & 1  & .423\\

Sig. (two-tailed)&  -& .195\\
N&11  & 11\\
\hline
\textbf{$AE$ \&$PS$ } &   & \\
Pearson r -& .423  & 1\\
Sig. (two-tailed)& &<.195\\
N&11  & 11\\
\hline

\end{tabular}
\end{table*}

On the other hand, Table \ref{tab:mytableccv} shows the correlation coefficients  between $AE$  and $AE$ and ($AE$ \& $PS$ for educators. As is shown in Table  \ref{tab:mytableccv}, the Pearson correlation, r between the variables  is $(r = 0.423, p < 0.05)$, which indicates a moderate positive correlation. This correlation is not statistically significant  $(p = 0.195)$. This suggested that there may not exist a strong relationship between the $AL$ and engagement among the educators. Also, the correlation coefficient for $AE$ \& $PS$ among the educators is 0.423 indicating a moderate correlation, which is not statistically significant $(p = 0.195)$, which suggests that there may not exist a significant relationship between $AL$ and engagement, and problem-solving skills among the educators.




The findings from this correlation analysis provide insights on the relationship between $AL$, and engagement and problem-solving among learners and educators. It has been observed that among the learners, there exist a strong positive correlation among $AL$, engagement and problem solving. This shows that as the learners actively engage in the learning process, their engagement levels and problems solving skills do increase too. However, that is not the case among the educator's perception; the correlation between $AL$, and engagement and problem-solving is not statistically significant-which shows that there is no strong relationship how they feel about these variables.


\section{Comparison with existing studies}
In this section a comparison that assesses the impact of the proposed approach with existing studies is conducted in the context of cybersecurity distance education. The motivation of exploring the comparative approach is to seek to examine and evaluate the similarities and differences that exist in the proposed/observations against the literature.  Using surveys administered  encompassing questionnaires shown in Tables \ref{tab:mytablevv} and \ref{tab:mytableww} respectively to a limited number of learners and educators with experience in cybersecurity distance courses.

A study of $VLabs$ environment as a method of educating cybersecurity researchers that has been highlighted  by ~\cite{nance2009virtual} shows a significant impact. Through this study, unique implementations are assessed from local to remote access. The findings show the need for having a significant amount of resources to create a virtual study environment. While this study remains relevant, it is presented from a generic perspective, given that it is not centered on student learning and engagement. Also, a personalized learning perspective in a Virtual Hands-on lab platform for Computer Science Education is able to identify individual student learning styles during lab sessions. This study went an extra mile to automatically assess, predict and monitor student learning and performance changes over time. While this approach was able to show a positive impact and efficiency on the student's, there was no measure on the degree of engagement ~\cite{deng2018personalized}. Another study that analyzed the students that had to learn in a container-based $VLabs$ for cybersecurity assessed the factors that influenced student learning, where indicators that were relevant to the student behavior on courses were assessed like estimated Effort (EE), Perceived Usefulness (PU), social influence, usefulness and intension of use. While this study was also relevant, it hardly had a focus on Active Engagement ~\cite{robles2019analyzing}. An assessment of student learning in the virtual environment in computer networks course reveals that in order for students to learn in a virtual laboratory, prior experience counts and the study distinguished the learning events in lectures and laboratories. The study is insightful as far as the effectiveness is concerned, however, it was viewed from the students' perspective ~\cite{wolf2009assessing}. Other relevant studies on the use of virtual labs include using $VLabs$ network testbed technology, which showed a positive effect on student knowledge, especially student thinking, when aligned to Bloom's taxonomy ~\cite{luse2021using}.  Also, the learners, support services can affect student engagement when online Learning environments are used as highlighted by ~\cite{he2019learner}. The findings suggest that when instructors use online learning Services (OLS) as indicators, it assists in understanding the learning processes among students.

\section{Discussions}

The findings from this study have provided insights on the positive impact of $VLabs$ on $AL$ and engagement in cybersecurity distance education. The survey results have revealed that both learners and educators had a favorable perception of the use of $VLabs$ in enhancing their understanding of cybersecurity concepts. The following sections discuss the key findings and their implications in more detail.

Based on the sample from the learners and the educators, this study has been able to examine the impact of $VLabs$ on $AL$ and engagement in Cybersecurity distance education. It was observed that a proportion of learners strongly agreed that using $VLabs$ in cybersecurity distance courses not only actively enabled them to engage in the course but also it had an effect in their problem-solving skills. A good proportion of learners agreed that $VLabs$ is a worthy approach as far as distance education is concerned. In addition, they found that to some extent, the $VLabs$ helped them to improve their practical skills in cybersecurity. However, a proportion of the learners were not certain but chose to stay neutral on this with some disagreeing. 

A percentage of respondents who agreed that $VLabs$ had a positive impact on their engagement and interest in the cybersecurity distance course is noteworthy. This suggests that among learners the $VLabs$ provide an effective means  to actively engage with the course material, problem-solving strategies resulting in continuous active engagement. The hands-on nature of $VLabs$ allow learners to apply theoretical knowledge in practical scenarios, fostering a deeper understanding of cybersecurity principles. It has been observed also that the positive impact on engagement is  important in distance education settings where learners may face challenges of continuous interaction with the course content.

 It has also been observed that $VLabs$ facilitates active engagement in problem-solving among learners. This is because, a significant proportion of respondents acknowledged that $VLabs$ had helped them actively engage in problem-solving activities, discussions, and hands-on exercises. This outcome consolidates the notion that $VLabs$ provide a a suitable platform and environment for learners  critically think and be equiped with problem-solving capabilities. 

It is worth to note that although a small percentage of respondents had a neutral   perception of the impact of $VLabs$, it was still seen that a good proportion considered the use of $VLabs$ to be essential. A number of factors that could lead to this could be attributed to lack of prior experiences or customed learning preferences or technical challenges. It is the authors opinion that  providing guidance and resources for learners to navigate and utilize $VLabs$ effectively and instructors role could overcome the prevailing concerns. 

Literature has also shown the need for active engagement for learners when utilizing remote resources. One of the key aspects that stimulated student engagement is the course materials, which act as a determinant on many learning outcomes. Since both the learners and educators had both interacted with $VLabs$ environment, it is the author's opinion that it could be important for the educators to initiate key strategies that could ensure that there is continuous engagement among the learners.

Consequently, correlation results show that the learner's perception on $VLabs$ engagement was significantly high coupled with problem-solving abilities when compared to the perception of the educators. The most ideal or probable reason for this can be attributed to the respondents having previous experience with $VLabs$ environment or respondents that are mainly from Computer Science discipline. 

It is also worth noting that the suggestions in this study are also aligned to the ICAP framework ~\cite{chi2014icap}, as mentioned in Section II of this paper, which shows an emphasis on how $AL$ and student engagement  promotes student learning. These claims can be supported by the scores shown from the outcome of this study.
\section{Limitations and Future Studies}

While important  insights have been pointed out in this study , it is worth acknowledging that there exist limitations and and it is also worth considering the avenues for future studies. The  sections that follow discuss the limitations that were encountered during the study and a suggestion of future direction is given. 

One limitation of this study is the use of relatively a small sample size of participants especially for the educators. Even though  efforts were made to have a diverse range of learners and educators who had experienced  cybersecurity distance courses leveraging $VLabs$, the sample size do not fully represent the entire population. For future studies it could be beneficial using  larger sample sizes and wider participant recruitment in order to enhance the generalizability of our findings.

There is also a  limitation relying on the reported data from survey responses. While the author agrees that surveys provide a convenient means of gathering  data, they could have potential biases.  Future studies could incorporate additional and relevant research methods, in order to gain a more harmonized and comprehensive understanding of the impact of $VLabs$ on $AL$ and engagement in cybersecurity distance education.

 While  it is important to note that the outcomes that that were expressed by participants are valuable, future research is projected to incorporate examination assessments in order to  objectively measure whether the effectiveness of $VLabs$ in improving learning outcomes in cybersecurity distance education can have an influende in pedagogical practices.

\section{ Conclusions}
The aim of this study was  to investigate the impact of $VLabs$  on $AL$ and engagement in cybersecurity distance education. From a survey of learners and educators that had experience in cybersecurity distance education that utilized $VLabs$, a survey was conducted and important insights have been reported. The findings of this study have shown that  $VLabs$ are viewed positively by both learners and educators, as they enhance active engagement and problem solving strategies. Furthermore, the outcome has shown that $VLabs$ have the potential to enhance $AL$ in cybersecurity distance education.

\section*{Declaration}
Data availability used in this research is available on request and the author has no competing interests.

\bibliographystyle{unsrtnat}
\bibliography{references}  






\end{document}